\documentclass[sigconf]{acmart}

\AtBeginDocument{%
  \providecommand\BibTeX{{%
    \normalfont B\kern-0.5em{\scshape i\kern-0.25em b}\kern-0.8em\TeX}}}

\setcopyright{acmcopyright}
\copyrightyear{2021}
\acmYear{2021}
\acmDOI{10.1145/1122445.1122456}

\acmBooktitle{}
\acmPrice{15.00}
\acmISBN{978-1-4503-XXXX-X/18/06}

\newcommand{\algname}[1] {{\fontfamily{cmtt}\selectfont {#1}}}
\usepackage{dsfont}

\usepackage{todonotes} 

\usepackage{caption}
\usepackage{subcaption}
\usepackage{enumerate}
\usepackage{multirow}
\usepackage{amsmath}

\begin{document}

\title{Unbiased Cascade Bandits: Mitigating Exposure Bias in Online Learning to Rank Recommendation}

\author{Masoud Mansoury}
\affiliation{
  \institution{University of Amsterdam}
  \city{Amsterdam}
  \country{the Netherlands}
}
\email{m.mansoury@uva.nl}

\author{Himan Abdollahpouri}
\affiliation{
    \institution{Northwestern University}
    \city{Evanston}
    \country{USA}
}
\email{himan.abdollahpouri@northwestern.edu}

\author{Bamshad Mobasher}
\affiliation{%
  \institution{DePaul University}
  \city{Chicago}
  \country{USA}}
\email{mobasher@cs.depaul.edu}

\author{Mykola Pechenizkiy}
\affiliation{%
 \institution{Eindhoven University of Technology}
 \city{Eindhoven}
 \country{the Netherlands}}
\email{m.pechenizkiy@tue.nl}

\author{Robin Burke}
\affiliation{%
  \institution{University of Colorado Boulder}
  \city{Boulder}
  \country{USA}}
\email{robin.burke@colorado.edu}

\author{Milad Sabouri}
\affiliation{%
  \institution{DePaul University}
  \city{Chicago}
  \country{USA}}
\email{msabouri@depaul.edu}

\renewcommand{\shortauthors}{Mansoury, et al.}

\begin{abstract}
  Exposure bias is a well-known issue in recommender systems where items and suppliers are not equally represented in the recommendation results. This is especially problematic when bias is amplified over time as a few popular items are repeatedly over-represented in recommendation lists. This phenomenon can be viewed as a recommendation \textit{feedback loop}: the system repeatedly recommends certain items at different time points and interactions of users with those items will amplify bias towards those items over time. This issue has been extensively studied in the literature on model-based or neighborhood-based recommendation algorithms, but less work has been done on online recommendation models such as those based on multi-armed Bandit algorithms. In this paper, we study exposure bias in a class of well-known bandit algorithms known as \textit{Linear Cascade Bandits}. We analyze these algorithms on their ability to handle exposure bias and provide a fair representation for items and suppliers in the recommendation results. Our analysis reveals that these algorithms fail to treat items and suppliers fairly and do not sufficiently explore the item space for each user. To mitigate this bias, we propose a \textit{discounting factor} and incorporate it into these algorithms that controls the exposure of items at each time step. To show the effectiveness of the proposed discounting factor on mitigating exposure bias, we perform experiments on two datasets using three cascading bandit algorithms and our experimental results show that the proposed method improves the exposure fairness for items and suppliers. 
  
  
  
\end{abstract}

\ccsdesc[500]{Information systems~Users and interactive retrieval}
\ccsdesc[500]{Information systems~Recommender systems}

\keywords{recommender system, cascading bandit, exposure bias, feedback loop}

\maketitle
\pagestyle{plain}

\section{Introduction}


Exposure bias in recommender systems refers to the fact that some items and suppliers are over-represented in recommendation results, while other items and suppliers are not adequately represented \cite{abdollahpouri2020multi,chen2020bias,singh2018fairness}. Due to the interactive nature of recommendation systems, this bias can be amplified over time as users interact with the recommended items at each time and their interactions would be used as input for recommendation algorithm in the subsequent times \cite{mansoury2020feedback}.
In this way, recommendation system and users form a feedback loop \cite{sinha2016deconvolving,jiang2019degenerate}. Various research efforts have studied the exposure bias in offline recommendation algorithms \cite{sun2019debiasing,mansoury2020feedback,mansoury2021graph}, but less work has been done on interactive recommendation algorithms such as online learning to rank algorithms where the system iteratively learns from users' interaction feedback by balancing exploration versus exploitation \cite{barraza2017exploration}.


Multi-armed bandit (MAB) algorithms are often used to systematically balance exploitation and exploration. Simply put, given $L$ arms as the items in the system, the learning agent pulls $K$ arms at a time to form the recommendation list and tries to maximize the reward based on the utility associated with items selected by the user. There are different MAB algorithms for managing the exploration and exploitation trade-off \cite{sutton2018reinforcement}. For example, 
$\epsilon$-Greedy MAB algorithms always exploit the best items that have been explored so far with the probability of $1-\epsilon$, and explores other items with probability of $\epsilon$.
Upper Confidence Bound (UCB) algorithms approach the problem differently by making their selections based on how uncertain the algorithm is about a given selection. 

Recently, a variant of MAB algorithms known as \textit{Cascading Bandits} \cite{kveton2015cascading,kveton2015combinatorial,li2016contextual,li2020cascading,zong2016cascading,hiranandani2020cascading} have been proposed that employ UCB for their operations and show superior performance in comparison to other MAB algorithms. We focus our analysis on this class of MAB algorithms. Besides their strength in increasing the click-through rate, these algorithms are also able to address \textit{position bias} \cite{craswell2008experimental,chuklin2015click} and improve the topical diversity of the recommendations. We specifically focus on three cascading bandit algorithms: \algname{CascadeLSB} \cite{hiranandani2020cascading}, \algname{CascadeLinUCB} \cite{zong2016cascading}, and \algname{CascadeHybrid} \cite{li2020cascading}.

Our goal in this work is to investigate the degree to which these cascading bandit algorithms are vulnerable to exposure bias. We investigate the extent to which these algorithms successfully explore the full list of available items and how fairly they recommend different items or suppliers. In other words, we determine how successful they are in terms of 1) giving enough choice to different users over time (user-level exploration) and 2) giving a fair exposure to different items or suppliers over the course of interactions with the users for a given period of time. We observe that these algorithms, despite their high emphasize on exploration, still fail to fully explore the available items and have high exposure bias both for items and suppliers. To address these issues, we propose a simple approach by discounting the items that have been recommended too frequently in the past by the algorithm. Experimental results using two datasets show that our approach can significantly improve the exposure and fairness of the recommendations while keeping the relevance of the recommendations as high as possible. 

\section{Background} \label{background}

In this section, we provide some background on the notion of exposure bias \cite{singh2018fairness} and summarize the key aspects of cascading bandit algorithms \cite{zong2016cascading,hiranandani2020cascading,li2020cascading}. 

\subsection{Exposure bias}

The concept of exposure bias has been studied by many researchers and different approaches are proposed to address this issue \cite{singh2018fairness,geyik2019fairness,mehrotra2018towards,mansoury2020fairmatch,abdollahpouri2020connection}. Singh and Joachims in \cite{singh2018fairness} discussed how exposure bias can lead to unfair treatment of different groups of users in ranking systems. They proposed a general framework for addressing the exposure bias when maximizing the utility for users in generating the ranked results under various fairness constraints.   

In domains like job recommendation where the task is recommending people (job seekers or qualified candidates), besides giving a fair exposure to each item (applicants), the solutions need to also take into account the fair exposure for people belonging to different sensitive groups (e.g. male versus female). In this line of research, Zehlike et al. \cite{zehlike2017fa} proposed a re-ranking algorithm to improve the ranked group fairness in recommendations. Their approach optimizes to achieve a predefined exposure for the minority group as much as possible, while maintaining the accuracy of the system. In another work, Geyik et al. in \cite{geyik2019fairness} explored the exposure bias in LinkedIn Talent Search where the distribution of applicants belong to different groups of sensitive attributes in recommendation lists do not follow the distribution of applicants' group in the initial search results. They showed that applicants belonging to the protected group are often under-recommended. To address this issue, the authors proposed an algorithm to achieve the desired distribution of applicants' group with respect to sensitive attributes in top-n recommendation results.

Some research efforts have tried to mitigate exposure bias by improving \textit{aggregate diversity} \cite{adomavicius2011improving,adomavicius2011maximizing,mansoury2020fairmatch} seeking to increase the number of unique items in the recommendation lists. Antikacioglu and Ravi in \cite{antikacioglu2017} proposed the idea of aggregate diversity maximization using a maximum flow approach. They showed that the minimum-cost network flow method can be efficiently used for finding recommendation subgraphs that optimizes the aggregate diversity. 
Mansoury et al. in \cite{mansoury2020fairmatch} proposed a graph-based algorithm that finds high quality items that have low visibility in the recommendation lists by iteratively solving the maximum flow problem on the recommendation graph.

In addressing the exposure bias for suppliers, Abdollahpouri in \cite{abdollahpouri2020popularity} analyzed the \textit{popularity bias} as the main source of exposure bias from the perspective of multiple stakeholders and proposed the idea of popularity calibration for mitigating this bias. Mehrotra et al. \cite{mehrotra2018towards} investigated the trade-off between the relevance of recommendations for users and supplier fairness, and their impacts on users' satisfaction. 
To determine the supplier fairness in recommendation list, first, suppliers are grouped into several bins based on their popularity in rating data and then the supplier fairness of a recommendation list is measured as how diverse the list is in terms of covering different supplier popularity bins.

\subsection{Cascading bandit models}

In a cascading bandit \cite{hiranandani2020cascading,li2020cascading,zong2016cascading}, the learning agent interacts with users by delivering the recommendations to them and receiving feedback. In these algorithms the \textit{Cascade model} \cite{craswell2008experimental,chuklin2015click} is used for modeling user's feedback behavior. The system presents the user with a list of recommended items. The user examines each recommended item one by one from the first position to the last, selects the first attractive item, and does not examine the rest of the items. This way, the items above the selected item are considered unattractive, the selected item is considered attractive. The rest of the items are considered as unobserved (neither attractive nor unattractive). Cascade model is effective in addressing well-known \textit{position bias} \cite{collins2018position,hofmann2014effects} where lower ranked items in the recommendation list are less likely to be clicked than the higher ranked items. By not considering the lower ranked (unobserved) items as unattractive, the model can still give them a chance to appear in the higher positions in the recommendations lists later interaction rounds.

The cascading bandit algorithms compute the probability of the target user liking a target item called \textit{attraction probability} using known item features and unknown user preferences in each iteration $t$. Since user preference is unknown and needs to be learned by interacting with the user, the attraction probability is estimated by solving a ridge regression problem over past observations on item features as independent variables and their attraction probabilities as dependent variable in $t-1$ time steps. Due to the uncertainty in estimating user preference toward items, cascading bandit algorithms employ UCB approach to model this uncertainty by computing an upper bound for the expected weight of each item. Given weights computed for each item, $K$ items with the largest estimated weights are returned to form the recommendation list $\mathcal{A}_t$. In the next step, given recommendation list $\mathcal{A}_t$, the agent receives feedback according to the cascade model and finally, for each examined item, the agent updates the model parameters.

The reward of the agent at each iteration is computed by $f(\mathcal{A},w)=1-\prod_{k=1}^K(1-w(a_k))$ which is 1 if user clicks on an item and 0 otherwise. Then, the performance of the agent's policy is evaluated by its \textit{n-step regret}:
\begin{equation}\label{eq_regret}
    R(n)=\sum_{t=1}^n \mathbb{E}[f(A^*,w_t)-f(\mathcal{A},w_t)]    
\end{equation}
\noindent where $A^*$ is the optimal list of items that maximizes the reward at any time $t$. Cascading bandits proposed in \cite{zong2016cascading,hiranandani2020cascading,li2020cascading} generally follow the above process for their operations and only differ in how they define the item features. 


\subsubsection{\algname{CascadeLSB}}
Yue and Guestrin in \cite{yue2011linear} proposed the notion of \textit{topic coverage} of an item to describe the probability that the item covers each topic. Similarly, the topic coverage of a list is defined as the probability of the items in the list covering each topic. Assuming one or more topics are associated with each item and topic coverage for each item, the feature vector for each item is defined as the gain in topic coverage by adding that item to the recommendation list created so far. Therefore, the feature vector for each item is dependent to the items already added to the list. If the target item is diverse in terms of topics with the items previously added to the recommendation list, the target item will have higher chance to be added to the list. Hiranandani et. al. in \cite{hiranandani2020cascading} adapted this algorithm by considering Cascade model to address the position bias and proposed \algname{CascadeLSB}.
 
\subsubsection{\algname{CascadeLinUCB}}
Wen et al. in \cite{wen2015efficient} proposed this algorithm that uses item features derived from the user-item interaction data and, unlike \algname{CascadeLSB} that only seeks to generate diversified recommendation, this algorithm aims at recommending items that are relevant to the user’s preferences. Various ways can be considered for deriving item features. One way proposed in \cite{li2020cascading} that we also used for our experiment is performing singular-value decomposition (SVD) on user-item interaction data. Zong et. al. in \cite{zong2016cascading} adapted this algorithm by considering Cascade model for interpreting user feedback behavior and proposed \algname{CascadeLinUCB}.

\subsubsection{\algname{CascadeHybrid}}
Li et. al. in \cite{li2020cascading} proposed \algname{CascadeHybrid} by combining \algname{CascadeLSB} and \algname{CascadeLinUCB} for generating recommendations that are both relevant to the users' preferences and diverse in terms of topics. 

\section{Bias analysis}

In this section, we analyze the cascading bandits described in section \ref{background} in terms of their ability in representing different items in recommendation lists over time. 

\subsection{Methodology}

For the experiments, we follow the experimental setting and data pre-processing used in \cite{li2020cascading}. We perform our experiments on two publicly-available datasets: MovieLens \cite{harper2015movielens} and Last.fm \cite{schedl2016lfm}. 

\subsubsection{Data pre-processing}

On MovieLens dataset, we extract 1K most active users and 1K most rated items from the user-item interaction data. Then, we convert the ratings to binary values where rating 5 is converted to 1 and others to 0. For our analysis on supplier-side exposure bias, we consider the movie-maker of each movie as the supplier. Since this dataset does not originally contain information about the movie-makers, we used the API provided by OMDB website\footnote{http://www.omdbapi.com/} to extract the information about movie-makers associated with different movies. Overall, there are 92,114 interactions between users and items, 18 genres associated to the items, and 513 suppliers (i.e. movie-makers) in this sample.

On Last.fm dataset, we extract 2K most active users and 2K most rated items from the user-item interaction data. To binarize the data, for items interacted more than 50 times by a user, we assign rating 1 and 0 to the others. Since the data does not have genre information, we extracted genre information using API on TheAudioDB\footnote{https://www.theaudiodb.com/}. Overall, there are 137,587 interactions between users and items, 88 genres associated to the items, and 967 suppliers (i.e. artists). 

\subsubsection{Simulation}

For simulating the interaction between learning agent and users, following the setting in \cite{li2020cascading,hiranandani2020cascading}, we randomly divide users profile into 50\% as training set and 50\% as test set. Training set is used for computing attraction probability of each item and generating the recommendation list to each user. Test set is used for modeling user feedback on recommendation list and generating the optimal recommendation list for evaluating the performance of the model. We perform the experiments for $n=10k$ steps and recommendations are generated for 100 randomly selected users. 

\subsubsection{Evaluation metrics}

We use the following metrics for measuring the exposure bias and the performance of recommendation results:

\begin{itemize}
    \item \textbf{n-step regret:} Cumulative difference between the optimal and agent's rewards computed by equation \ref{eq_regret}.
    \item \textbf{Item Coverage (IC):} The fraction of items recommended at least once to all users over $n$ interactions.
    \item \textbf{Supplier Coverage (SC):} The fraction of suppliers recommended at least once to all users over $n$ interactions.
    \item \textbf{Average User-level Item Coverage (UIC):} Average item coverage for each user.
    \item \textbf{Gini Index (G):} Uniformity measure of the frequency distribution of recommended items. Gini Index is in $[0,1]$ where 0 indicates uniform distribution (fair outcome) and 1 indicates skewed distribution (unfair outcome).
\end{itemize}

\begin{figure*}[!t]
    \centering
    \begin{subfigure}[b]{0.98\textwidth}
        \includegraphics[width=\textwidth]{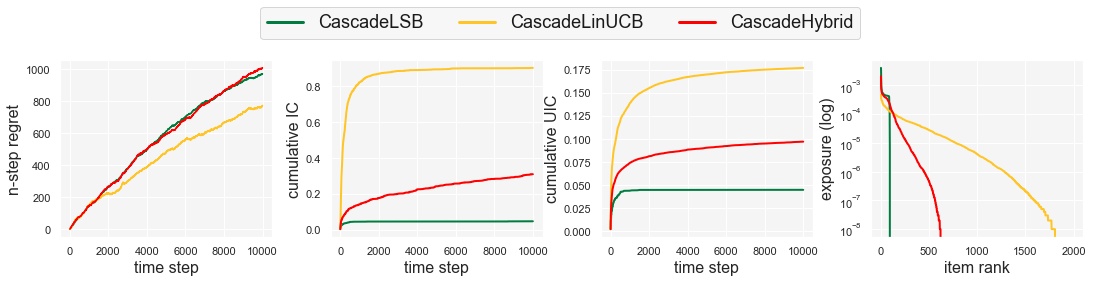}
        \caption{Last.fm} \label{bias_lf}
    \end{subfigure}
    \begin{subfigure}[b]{0.98\textwidth}
        \includegraphics[width=\textwidth]{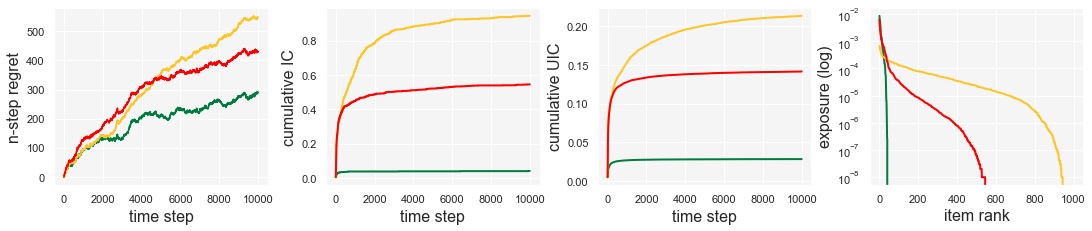}
        \caption{MovieLens} \label{bias_ml}
    \end{subfigure}%
\caption{The performance of cascading bandit algorithms in addressing exposure bias for $c=1$.} \label{bias}
\end{figure*}

\subsection{Experimental Results}

Figure \ref{bias} compares these algorithms from four different perspectives on Last.fm (Figure \ref{bias_lf}) and MovieLens (Figure \ref{bias_ml}) datasets and shows how they handle exposure bias in recommendation results.

On Last.fm, in terms of n-step regret, \algname{CascadeLinUCB} achieved better performance (lower n-step regret), while on MovieLens, \algname{CascadeLSB} outperformed other algorithms. This is mainly because \algname{CascadeLSB} behaved differently on each dataset. Item-topic data on Last.fm is much sparser than the item-topic data on MovieLens. Since \algname{CascadeLSB} only uses topic information, highly sparse topic data can lead to poor performance. Also, the performance of \algname{CascadeHybrid} is between \algname{CascadeLinUCB} and \algname{CascadeLSB} which is expected as it is a combination of those algorithms.

Cumulative item coverage simply shows how each algorithm explores item space over time by recommending new items to users. A flat line in the plot indicates no new item was added to the recommendation lists over time. On both datasets, we can see that all algorithms explore the item space more in the early iterations and then they focus on exploitation in the later iterations. \algname{CascadeLSB} stops exploring the item space after few iterations which leads to not covering sufficient items in the recommendation lists. Also, the plot shows that even running the algorithm for more iterations does not improve the item coverage. \algname{CascadeLinUCB} is performing well by covering majority of the items in the recommendation lists and the ascending slope of the plot suggests that it may further improve item coverage if we run the algorithm for a longer time period. Although \algname{CascadeLinUCB} showed performing well so far, our next analysis reveals its weaknesses.

Cumulative user-level item coverage shows the item coverage for each user and how many new items are exposed to each user at each iteration. Analogous to what we observed for cumulative item coverage, \algname{CascadeLinUCB} outperformed other algorithms, \algname{CascadeLSB} is the weakest by recommending only a small fraction of the items, and \algname{CascadeHybrid} is performing in between. An interesting pattern in these results is that even though \algname{CascadeLinUCB} performs well in terms of item coverage by covering more than 80\% of the items on both datasets in the whole recommendation lists for all users, its user-level item coverage is very low by covering only less than 18\% of the items on Last.fm dataset and less than 22\% of the items on MovieLens dataset. This shows that these algorithms are unable to explore sufficiently the item space for an individual user and clearly they recommend the same items repeatedly to that user over time. This issue is more severe for \algname{CascadeLSB} on both datasets and \algname{CascaseHybrid} on MovieLens which the algorithms even stop exploring the item space and recommending new items to each user.

\captionsetup[table]{skip=4pt}
\begin{table*}[t]
\centering
\setlength{\tabcolsep}{3pt}
\captionof{table}{Comparison of original and unbiased cascading bandit algorithms in terms of item-side and supplier-side fairness for $c=1$. The bolded entries show the best values and the underlined entries show the statistically significant change with $p<0.01$.} \label{tab:res}
\begin{tabular}{lrrrrrrrrr}
\toprule
 \multirow{2}{*}{algorithms} & \multicolumn{4}{c}{Last.fm} & & \multicolumn{4}{c}{MovieLens} \\\cline{2-5}\cline{7-10}
 
 & $IC$ & $SC$ & $UIC$ & $G$ & & $IC$ & $SC$ & $UIC$ & $G$ \\
 \bottomrule
 
 \algname{CascadeLSB}  &  
     4.6\% & 9.1\% & 4.5\% & 0.964 && 
     4.1\% & 5.7\% & 2.8\% & 0.987 \\
 \algname{CascadeLinUCB} &
    90.4\% & 96.3\% & 17.7\% & 0.759 && 
    94.4\% & 96.5\% & 21.3\% & 0.639 \\
 \algname{CascadeHybrid} &
    30.9\%& 44.7\% & 9.7\% & 0.937 && 
    54.2\% & 65.7\% & 14.1\% & 0.954 \\
 \hline
 \algname{UnbiasedCascadeLSB}  &  
    \textbf{5.7\%} & \textbf{10.5\%} & \underline{\textbf{5\%}} & \textbf{0.962} && 
    \textbf{5.2\%} & \textbf{7.4\%} & \underline{\textbf{3.7\%}} & \textbf{0.985} \\
 \algname{UnbiasedCascadeLinUCB} &
    \underline{\textbf{96.7\%}} & \underline{\textbf{99.2\%}} & \underline{\textbf{35.4\%}} & \underline{\textbf{0.630}} && 
    \underline{\textbf{97.5\%}} & \underline{\textbf{98.6\%}} & \underline{\textbf{35\%}} & \underline{\textbf{0.550}} \\
 \algname{UnbiasedCascadeHybrid} &
    \underline{\textbf{36.2\%}} & \underline{\textbf{49.3\%}} & \underline{\textbf{11\%}} & \textbf{0.929} && 
    \underline{\textbf{60.8\%}} & \underline{\textbf{72.1\%}} & \underline{\textbf{18.1\%}} & \textbf{0.942} \\
 
 \bottomrule
\end{tabular}
\end{table*}

Finally, \textit{exposure} (the rightmost plot) shows the distribution of the recommendation frequency for different items (number of times each item appeared in the recommendation lists). Since the distribution for all algorithms are extremely long-tailed and their plots overlap, for the presentation purposes, we performed log-transformation on the values. These results also confirm that all three algorithms extremely suffer from exposure bias by not fairly representing the items in the recommendation lists. The long-tail distribution of the recommendation frequency for the items shows that few items frequently appeared in the recommendation lists (over-recommendation), while majority of the items rarely appeared in the recommendation lists (under-recommendation). 

Overall, the results in Figure \ref{bias} reveal that the cascading bandit algorithms described in section \ref{background} suffer from exposure bias where items are not equally represented in the recommendation lists. In the following sections, we show that how this unfair treatment of items can lead to supplier-side unfairness. Therefore, a solution is needed to address the issue of exposure bias in cascading bandits. In the next section, we propose our solution for tackling exposure bias for items and suppliers, and we show how it can improve exposure fairness. 
 
\section{Mitigating Bias in Cascading Bandits}

In this section we integrate our proposed discount factor into the cascading bandit algorithms and empirically examine the degree to which the extended algorithms can mitigate exposure bias.

\subsection{Cascading Bandits with Exposure Discount}

The utility function in bandit algorithms described in section \ref{background} are in the following form: 
\begin{equation}
    utility_t(i) = AttractionProb_t(i) + c \times UpperBound_t(i)
\end{equation}
\noindent where $c$ controls the degree of exploration. Higher value for $c$ allows the algorithm to explore more which, as a result, will degrade the performance of the algorithm in terms of n-step regret. This utility function does not differentiate between frequently and rarely exposed items and assigns the same degree of exploration to all items. To better control the degree of exploration for items with different exposure level, we introduce a \textit{discounting factor} that computes the degree of exploration for each item according to its exposure until time $t$. Thus, the utility function would be as follows:

\begin{equation}\label{unbiased_utility}
    utility_t(i) = AttractionProb_t(i) + c \times (1-\frac{N_i}{t}) \times UpperBound_t(i)
\end{equation}

\noindent where $N_i$ is the number of times item $i$ is recommended in the past $t-1$ time steps for a particular user to avoid unnecessary repetitions. For items that are already highly exposed, this discounting factor decreases the degree of the exploration, while for rarely exposed items, it increases the degree of exploration to give them more chance to be recommended. For the rest of the paper, for each algorithm that uses equation \ref{unbiased_utility} for computing the utility, we name it the unbiased version of that algorithm: \algname{UnbiasedCascadeLSB}, \algname{UnbiasedCascadeLSB}, and \algname{UnbiasedCascadeHybrid}.

\subsection{Experimental results}

In this section, we evaluate our unbiased cascading bandits in terms of their ability in addressing item-side and supplier-side exposure fairness and compare them with the original cascading bandits described in section \ref{background}. 

\subsubsection{The Impact on Item and Supplier Fairness}

Table \ref{tab:res} presents the experimental results for original and unbiased cascading bandit algorithms for $c=1$ as it resulted in a better performance in terms of n-step regret and exposure fairness on original cascading bandits. The results show that on both datasets and across all metrics, our unbiased cascading bandits significantly outperform the original ones.

One important consideration for analyzing the fairness of an algorithm is how many unique items or suppliers are represented in the recommendation lists. In particular, in bandit algorithms where the algorithm is running for a long period of time, it is expected that more items or suppliers to have a chance to appear in the recommendation lists. This is measured by $IC$ and $SC$ in Table \ref{tab:res}.The proposed unbiased algorithms significantly improved the coverage for items and suppliers with respect to these two metrics as compared to the original algorithms. For instance, \algname{UnbiasedCascadeHybrid} yielded 36.2\% $IC$ and 49.3\% $SC$ on Last.fm compared to 30.9\% and 44.7\% for original algorithms, respectively, and 60.8\% $IC$ and 72.1\% $SC$ on MovieLens compared to 54.5\% and 65.7\% for original algorithms, respectively. 

Another important consideration is the coverage of items for each user. Low item coverage for a user means that few distinct items are repeatedly recommended to that user in different time steps. This will have negative impacts on users experience as users are not exposed to new items and the algorithm may not be able to properly learn various preferences. This is evaluated by $UIC$ in Table \ref{tab:res}. As shown,
the proposed unbiased cascading bandits achieved fairer results in terms of $UIC$ compared to the original algorithms. For instance, \algname{UnbiasedCascadeLinUCB} yielded 35.4\% and 35\% $UIC$ on Last.fm and MovieLens datasets compared to 17.7\% and 21.3\% $UIC$ for \algname{CascadeLinUCB}, respectively.


Finally, fair distribution of recommended items is another important factor that shows how the system equitably treats items when generating recommendation lists for users. This is shown by $G$ in Table \ref{tab:res}. Again, the proposed unbiased bandits outperformed the original algorithms in terms of $G$. For instance, \algname{UnbiasedCascadeLinUCB} outperformed \algname{CascadeLinUCB} by achieving $G$ of 0.630 compared to 0.759 on Last.fm and 0.550 compared to 0.639 on MovieLens.

\begin{figure*}[!t]
    \centering
    \begin{subfigure}[b]{0.98\textwidth}
        \includegraphics[width=\textwidth]{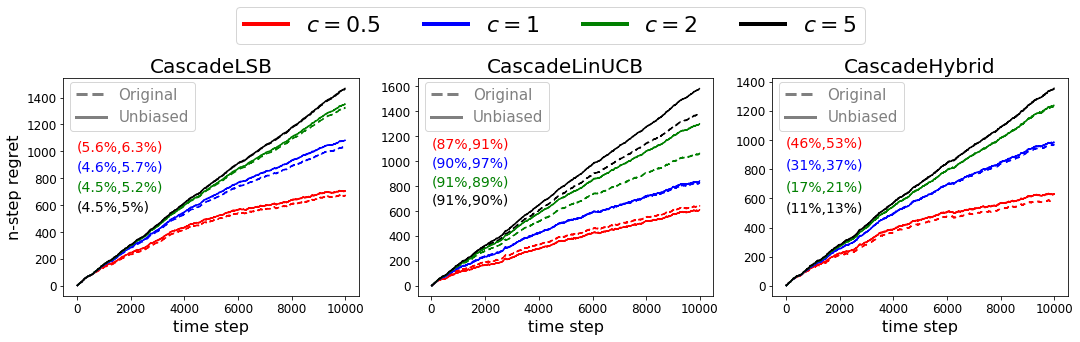}
        \caption{Last.fm} \label{regret_lf}
    \end{subfigure}
    \begin{subfigure}[b]{0.98\textwidth}
        \includegraphics[width=\textwidth]{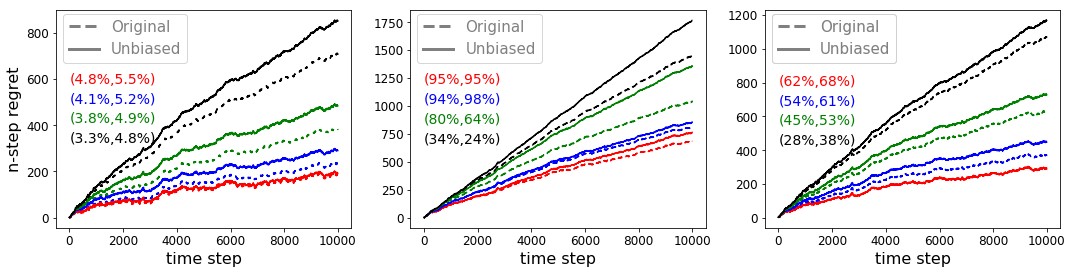}
        \caption{MovieLens} \label{regret_ml}
    \end{subfigure}%
\caption{n-step regret for the original (dashed) and unbiased (solid) algorithms for different values of $c$. Lower regret indicates that more clicks are received by the algorithm during the online learning. The numbers in the tuple show the item coverage for the corresponding $c$ values. The first and second numbers are the item coverage for original and unbiased algorithm, respectively.} \label{regret}
\end{figure*}

\subsubsection{The Effect of Varying Exploration Coefficient}

Coefficient $c$ in the utility function controls the degree of exploration in the cascading bandits. This means that we should be able to explore more by increasing $c$. Thus, an important question is: \textit{can we tune coefficient $c$ to achieve the highest possible exposure fairness for items and suppliers without the proposed discounting factor in equation \ref{unbiased_utility}?} In this section, we empirically answer this question by performing experiments with different values of $c$ and the experimental results show that tuning $c$ does not necessarily improve the exposure fairness for items and suppliers.

Figure \ref{regret} shows the performance of original and unbiased cascading bandit algorithms for $c \in \{0.5,1,2,5\}$ in terms of n-step regret. The numbers shown as a tuple in each plot show the item coverage ($IC$) where the first and second numbers indicate the item coverage of original and unbiased algorithm, respectively. Looking at the plots for original algorithms (dashed line) reveals that when we increase $c$, the n-step regret increases over time (lower performance). However, higher $c$ (higher weight for exploration) does not necessarily improve $IC$. The same pattern is also observed for other metrics. This shows that even tuning $c$ may not necessarily increase the exposure of items (and suppliers). 


Comparing the performance of original and unbiased bandits in Figure \ref{regret} shows that for $c=0.5$ (red line), both original and unbiased bandits achieved almost the same n-step regret (even better performance by \algname{UnbiasedCascadeLinUCB} on Last.fm), but the item coverage achieved by our unbiased bandit is significantly higher than the original one. Also, the same pattern can be observed for $c=1$ in most cases. For $c \in \{2,5\}$, although the unbiased cascading bandits achieved higher n-step regret than original bandit, the performance for both versions is not practical as both n-step regret and item coverage are worse than the ones achieved from $c \in \{0.5,1\}$. In practice, the hyperparameters are tuned to achieve the best performance. In our experiments, $c \in \{0.5,1\}$ yielded the best performance in terms of both n-step regret and item coverage in all cases, therefore, we concentrate our analysis on the results by $c \in \{0.5,1\}$.

On Last.fm dataset in Figure \ref{regret_lf}, the proposed unbiased cascading bandits resulted in the highest item coverage with insignificant change in n-step regret as 6.3\% ($c=0.5$) for \algname{UnbiasedCascadeLSB}, 97\% ($c=1$) for \algname{UnbiasedCascadeLinUCB}, and 53\% ($c=0.5$) for \algname{UnbiasedCascadeHybrid}. On MovieLens, the same pattern is observed with the highest item coverage as 5.5\% ($c=0.5$), 98\% ($c=1$), and 68\% ($c=0.5$), respectively. Also, another interesting pattern is that for all $c$ values on both datasets, the proposed unbiased cascading bandits yielded higher item coverage than the original ones, except for \algname{CascadeLinUCB} for $c \in \{2,5\}$. All these results show that the proposed discounting factor is effective in addressing exposure bias and is able to significantly improve the exposure fairness of items and suppliers.

\section{Conclusion and Future Work}

In this paper, we studied exposure bias in three well-known cascading bandit algorithms. Our results showed that these algorithms do not fairly represent items and suppliers in the recommendation lists over time. To address this issue, we introduced a \textit{discounting factor} and incorporated it into the utility function of those algorithms. This discounting factor is able to adjust the degree of exploration for different items based on their exposure in previous time steps. Our experiments on two datasets showed that the proposed discounting factor significantly improved the exposure of items and suppliers with negligible loss on reward. 

In future work, following the idea proposed in \cite{li2016contextual}, we plan to investigate the effectiveness of discounting the feature vector of highly exposed items when updating the model parameters at each iteration, instead of incorporating the discounting factor into the utility function, for mitigating the exposure bias. Another interesting future work is taking into account the position of each item in the proposed discounting factor. This means that while trying to balance the exposure for different items, we also need to provide equal opportunity for different items to be shown on the top of the list. This is in particular important in the domains like job recommendation where people belonging to different sensitive groups must have chance to appear on top of the lists \cite{zehlike2017fa,geyik2019fairness}.  

We also plan to investigate the relationship between the popularity of the items in the input data and their exposure in the recommendation lists in bandit algorithms. This helps to better understand the factors leading to unfairness in these algorithms. Finally, as another future work, we are interested in reproducing the experiments conducted in this paper on other variants of MAB algorithms and analyzing those algorithms in handling exposure bias.

\bibliographystyle{ACM-Reference-Format}
\bibliography{ref}

\end{document}